\documentstyle[amssymb,aps,preprint]{revtex}

\begin{document}
\title{Intrasubband and Intersubband Electron Relaxation in Semiconductor Quantum
Wire Structures}
\author{Marcos R.S. Tavares and S. Das Sarma}
\address{Department of Physics, University of Maryland, College Park, MD 20742-4111,\\
USA}
\author{Guo-Qiang Hai}
\address{Instituto de F\'{i}sica de S\~{a}o Carlos, Universidade de S\~{a}o\\
Paulo, S\~{a}o Carlos, SP 13560-970, Brazil}
\maketitle

\begin{abstract}
We calculate the intersubband and intrasubband many-body inelastic Coulomb
scattering rates due to electron-electron interaction in two-subband
semiconductor quantum wire structures. We analyze our relaxation rates in
terms of contributions from inter- and intrasubband charge-density
excitations separately. We show that the intersubband (intrasubband)
charge-density excitations are primarily responsible for intersubband
(intrasubband) inelastic scattering. We identify the contributions to the
inelastic scattering rate coming from the emission of the single-particle
and the collective excitations individually. We obtain the lifetime of hot
electrons injected in each subband as a function of the total charge density
in the wire.
\end{abstract}

\pacs{73.61.r; 73.50.Gr; 72.10.Di.}

\section{Introduction}

\smallskip Semiconductor quantum wire structures, based mostly on
GaAs-AlGaAs systems, have been studied intensively for the last ten years as
systems of potential technological interest (e.g. quantum wire lasers), and
also because of their fundamental significance as examples of
quasi-one-dimensional (Q1D) electron liquids. Among the important research
milestones in semiconductor quantum wires are the observation \cite{1s} of
one dimensional plasmons via the inelastic light scattering spectroscopy and
the verification of the predicted acoustic linear plasma dispersion relation 
\cite{2s} in one dimension, the observation of pronounced one dimensional
Fermi edge singularities in the optical spectra \cite{3s}, the quantum wire
excitonic laser operation \cite{4s} and its theoretical understanding \cite
{5s}. With improving materials growth and nano-fabrication techniques one
expects a wide range of one dimensional experimental phenomena and projected
applications in semiconductor quantum wire systems. Many of the projected
applications such as ballistic electron transistors, quantum wire-based
infrared photo-detectors and lasers, quantum wire THZ oscillators and
modulators, will utilize fast carriers (injected or excited) in doped
quantum wires as the active device element. Effective control and
manipulation of these fast electrons in doped quantum wire systems are
therefore essential in the projected quantum wire opto-electronic
applications. One of the most crucial physical processes that will limit the
quantum wire opto-electronic applications is the relaxation of these fast
electrons. The main ultrafast mechanism controlling the relaxation process
is the electron-electron interaction, which is also a many-body process of
fundamental importance in electronic systems. In this paper, we develop a
many-body theory for the electron-electron interaction induced ultrafast
relaxation in semiconductor quantum wires with more than one quantized
subband occupied. We consider only the ultrafast electron-electron
interaction induced relaxation in this article, neglecting the weaker
electron-phonon Fr\"{o}hlich coupling. The electron-phonon coupling may be
considered to be approximately included in our theory by taking the
effective mass entering the theory to be the polaronic band mass (including
the electron-LO phonon interaction) and the background dielectric constant
to be the static low frequency lattice dielectric constant (rather than the
dynamic high frequency dielectric constant). We restrict ourselves to
Coulomb scattering because the fastest relaxation time scales are controlled
by the inter-electron Coulomb interaction.

We mention that semiconductor quantum wires in the strict one dimensional
limit with only one occupied subband are extremely difficult to fabricate.
Thus the typical experimental quantum wires would have a few occupied
subbands and scattering between these subbands effectively destroys their
strict one dimensionality. The work presented in this article takes a first
step toward developing a full many-body theory for quantum wires with many
occupied subbands by considering carefully the situation with two occupied
subbands and by analyzing the resultant relaxation rates in terms of
intrasubband (one dimensional) and intersubband (non-one dimensional)
scattering contributions. In addition, we calculate single-particle and
collective mode contributions to the relaxation rates separately. Our
calculations can be directly compared with experimentally measured
linewidths (e.g. the spectral width in tunneling measurements \cite{6s} or
in the femtosecond spectroscopy \cite{7s} ) or band broadenings and with
various relaxation rates entering device modelling considerations.

Intra- and intersubband relaxation of electrons in Q1D doped semiconductor
quantum wires are determined by their inelastic lifetime which is inversely
proportional to the inelastic Coulomb scattering rate. Due to the Coulomb
interaction, electrons in the quantum wires may be scattered and, as a
result, collective ('plasmons') and single-particle excitations are emitted.
Such lifetime calculations have earlier been carried out in 2D electrons
systems \cite{8s,9s}, and have been interpreted in terms of plasmon emission
processes. In contrast to 2D electrons gases, a gap shows up in the
intersubband single-particle excitation continuum in Q1D quantum wires with
two occupied subbands. \cite{2s,1,2,3} Furthermore, an extra intersubband
plasmon mode appears within such a gap. It was also shown previously that,
for a two-subband quantum wire, \cite{zheng} the intersubband
inelastic-scattering rates due to plasmon modes and single-particle
excitations do not exist if the intersubband coupling is neglected. But the
intrasubband inelastic-scattering rates were found consist of three
contributions: that coming from the emission of plasmon modes in the (i)
first and (ii) second subband; and the emission of a (iii) single-particle
excitation in the second subband. So far in the literature the intersubband
coupling in quantum wires has been considered irrelevant for electron
relaxation in the conduction band. However, as the second subband becomes
occupied, electrons in different subbands may interact strongly with each
other, and as a consequence, intersubband coupling should in general be
taken into account. In this paper we calculate the intra- and intersubband
inelastic-scattering rates of electrons in a two-subband quantum wires with
a small energy separation between the two subbands. We neglect the higher
lying subbands to reduce computational complications. A generalization of
our theory to many subbands is, in principle, possible. \smallskip Currently
there are no direct experimental observations of intersubband lifetimes in
multisubband quantum wires, but our calculations should be relevant to a
large number of projected applications.

We develop our theory for the inelastic Coulomb scattering treating the
dynamical screening of the Q1D electron system within the framework of the
random-phase approximation (RPA). The RPA has been shown to be an excellent
approximation in studying charge-density excitations in Q1D doped
semiconductors by virtue of the approximate vanishing of all vertex
corrections to the one-dimensional irreducible polarizability. \cite{2s,4}
In addition to the 1D intrasubband plasmons, the intersubband collective and
single-particle excitations in the Q1D system also provide relaxation
channels through which the hot electrons in the conduction band relax. We
show that the inelastic-scattering rate from the second to the first subband
can only occur through the emission of an intersubband plasmon with
single-particle excitations not participating at all in this intersubband
relaxation process, whereas emission of both collective and single-particle
excitations contributes to the inelastic scattering from the first to the
second subband.

\smallskip This paper is organized as follows. In Sec. II we describe our
theoretical approach. In Sec. III we present our numerical results for the
inelastic scattering rates in a two-subband quantum wire. We conclude with a
summary in Sec. IV.

\section{Model and Equations}

The single electronic states in our theory are calculated by considering a
two-dimensional system in the $xy$ plane subjected to an additional
confinement in the $y$-direction creating a GaAs/AlGaAs quantum wire in the $%
x$-direction. The confinement potential in the $y$-direction is taken to be
of a finite square well type of barrier height $V_{0}$ and well width $W$.
We assume the confinement potential creating the 2D confinement to be
sufficiently strong compared with the 1D confinement potential and assume
the 2D system to be ideal, i.e. of zero thickness in the third (z)
direction. The 1D subband energies $E_{n}$ and the wave functions $\phi
_{n}(y)$ are obtained from the numerical solution of the one-dimensional
Schr\"{o}dinger equation in the $y$-direction (the value of the electron
effective mass throughout this paper is $m^{*}=0.07m_{e}$). We restrict
ourselves to the case where $n=1,2$ and define $\omega _{0}=E_{2}-E_{1}$ as
being the intersubband energy gap between the two subbands. For a symmetric
confinement potential, the two lowest wave functions $\phi _{1}(y)$ and $%
\phi _{2}(y)$ are the usual symmetric and antisymmetric levels,
respectively. We consider throughout this paper the confinement potential
being of a barrier height $V_{0}$ = 100 meV and well width $W=500$ \AA ,
which leads to $\omega _{0}\simeq \,$5.37 meV. Then, the second subband
becomes populated at 1D electron density $N_{e}=6.3\times 10^{5}$ cm$^{-1}$.
As mentioned in the introduction, we neglect inelastic scattering due to
emission of phonons. Such a procedure is reasonable since the emission of an
LO phonon, for example in GaAs, requires the electron energy to be at least $%
\hbar \omega _{LO}\simeq 36$ meV which is much larger than the
characteristic inter-subband energy $\omega _{0}$ of our quantum wire. We
will restrict ourselves to situations where phonon emission processes are
not important. We take $\hbar =1$ throughout this paper unless stated
otherwise.

As we mentioned in the introduction, intra- and intersubband relaxation of
fast electrons in two-subband quantum wire structures can be studied in
determining their inelastic Coulomb scattering rate $\sigma _{nn^{\prime
}}(k)$, where $n,n^{\prime }=1,2$. Due to Coulomb interaction, these
electrons, initially in a subband $n$ with momentum $k$, can be scattered to
a subband $n^{\prime }$ with momentum $k^{\prime }$ through emission of both
plasmons and single-particle excitations. Within the so-called GW
approximation, Vinter \cite{vinter} originally showed that, at zero
temperature, the inelastic Coulomb scattering rate $\sigma _{nn^{\prime
}}(k) $ of electrons in multisubband structures can be obtained from the
imaginary part of the retarded electron self-energy neglecting higher order
vertex corrections. This approximation is extensively employed in
calculating electronic many-body effects and, in particular, has been used
in studying injected electron lifetimes in semiconductor quantum wire
structures in the strict one-dimensional limit (or equivalently in
determining the intrasubband inelastic scattering rate of electrons in the
first quantized subband), \cite{zheng,hwang91,BYK} as well as in coupled
parallel quantum wires.\cite{preprinosso} Within the GW approximation, the
multisubband inelastic scattering rate $\sigma _{nn^{\prime }}(k)$ of fast
electrons in quasi-one-dimensional quantum wires at zero temperature is
given by 
\[
\sigma _{nn^{\prime }}(k)=\frac{1}{2\pi }\int dq%
\mathop{\rm Im}%
\left\{ V_{nn^{\prime }n^{\prime }n}^{s}[q,\xi _{n^{\prime }}(k+q)-\xi
_{n}(k)]\right\} 
\]
\begin{equation}
\times \left\{ \theta \left( \xi _{n}(k)-\xi _{n^{\prime }}(k+q)\right)
-\theta \left( -\xi _{n^{\prime }}(k+q)\right) \right\} ,  \label{sigma}
\end{equation}
where $\theta \left( x\right) $ is the standard step function, $%
V_{nn^{\prime }n^{\prime }n}^{s}(q,\omega )$ the dynamically screened
electron-electron Coulomb potential with $q$ being the 1D wavevector and $%
\omega \,$ the mode frequency, and $\xi _{n}\left( k\right) =\hbar
^{2}k^{2}/2m^{*}+E_{n}-E_{F}$ the electron energy with respect to the Fermi
energy $E_{F}$. The screened Coulomb potential in Eq. (\ref{sigma}) is
related to the multisubband dielectric function $\varepsilon _{nn^{\prime
}mm^{\prime }}(q,\omega )$ and the bare electron-electron interaction
potential $V_{nn^{\prime }mm^{\prime }}(q)$ through the generalized RPA
equation \cite{mhbook} 
\begin{equation}
\sum_{ll^{\prime }=1,2}\epsilon _{ll^{\prime }nn^{\prime }}(q,\omega
)V_{ll^{\prime }mm^{\prime }}^{s}(q,\omega )=V_{nn^{\prime }mm^{\prime }}(q),
\label{diel}
\end{equation}
with $m,m^{\prime }=1,2$. The bare electron-electron potential $%
V_{nn^{\prime }mm^{\prime }}(q)$, which is the 2-particle matrix element of
1D Coulomb interaction in the $\phi _{n}(y)$ basis, is calculated by using
the numerical solution of the electron wavefunction $\phi _{n}(y)$. The
dielectric function 
\begin{equation}
\varepsilon _{nn^{\prime }mm^{\prime }}(q,\omega )=\delta _{nm}\delta
_{n^{\prime }m^{\prime }}-\Pi _{nn^{\prime }}(q,\omega )V_{nn^{\prime
}mm^{\prime }}(q)  \label{diel22}
\end{equation}
is calculated within the RPA, where 
\begin{equation}
\Pi _{nn^{\prime }}(q,\omega )=\frac{1}{\pi }\int dk\frac{n_{F}\left[ \xi
_{n}(k)\right] -n_{F}\left[ \xi _{n^{\prime }}(k+q)\right] }{\xi _{n}(k)-\xi
_{n^{\prime }}(k+q)+\omega }  \label{pol1}
\end{equation}
is the noninteracting irreducible polarizability function. Here, $n_{F}(E)$
is the Fermi distribution function. The polarizability $\Pi _{nn^{\prime
}}(q,\omega )$ characterizes the bare electron-hole bubble polarization
diagram and is written for the system free from any impurity scattering. The
impurity scattering effects can be introduced diagrammatically by including
impurity ladder diagrams in the electron Green's function. These diagrams
are responsible for level broadening, or equivalently, for a
phenomenological damping constant $\gamma =e^{2}/2m^{*}\mu $ mainly induced
by scattering of electrons due to impurity centers, with $\mu $ being the
carrier mobility in the sample. The exact expression for the polarizability
within this diagrammatic approach can be obtained by using a
particle-conserving approximation for arbitrary values of $q$ and $\omega $,
given by Mermin. \cite{Mermin} In the limit $\gamma \rightarrow 0$, the
Mermin's polarizability is found to be identical to Eq. (\ref{pol1}) with
the frequency $\omega ^{2}\rightarrow \omega (\omega +i\gamma )$. In this
paper, we take the impurity scattering induced broadening $\gamma $ as being
a very small phenomenological damping parameter which allows us working in
the limit $\gamma \rightarrow 0$. We are therefore restricting ourselves to
high mobility quantum wires with small level broadening.

According to Eq. (\ref{sigma}), the integral in $\sigma _{nn^{\prime }}(k)$
is performed only over the segment of the curve 
\begin{equation}
\omega _{k}^{nn^{\prime }}(q)=\xi _{n^{\prime }}(k+q)-\xi _{n}(k)
\label{omega11}
\end{equation}
which lies inside those regions where 
\begin{equation}
\theta \left[ -\omega _{k}^{nn^{\prime }}(q)\right] -\theta \left[ -\xi
_{n^{\prime }}(k+q)\right] \neq 0.  \label{teta}
\end{equation}
We need to consider, therefore, just that segment of $\omega
_{k}^{nn^{\prime }}(q)$ which lies in the region where the condition defined
in Eq. (\ref{teta}) is satisfied. The inelastic scattering rates vanish
outside these regions which means that the momentum and energy conservation
cannot be simultaneously obeyed for those values of ($k,n,n^{\prime },k+q$).
The inelastic scattering rate $\sigma _{nn^{\prime }}(k)$ is a non-vanishing
term if the segment $\omega _{k}^{nn^{\prime }}(q)\,$ either lies in the
continuum representing single-particle excitations or intercepts the lines
representing collective excitations (plasmons) in the $q$-$\omega $ plane.

\section{Numerical results and discussions}

\subsection{Collective and single-particle excitation modes}

In our quantum wires we consider a symmetric square well potential
characterizing the $y$-direction confinement. Due to this symmetry, the two
lowest wave functions $\phi _{1}(y)$ and $\phi _{2}(y)$ are symmetric and
antisymmetric functions of $y$, respectively. As a result, the bare
electron-electron Coulomb potential $V_{nn^{\prime }mm^{\prime }}(q)$
vanishes if $n+n^{\prime }+m+m^{\prime }$ is an odd number. Moreover 
\begin{equation}
V_{nn^{\prime }mm^{\prime }}(q)=V_{n^{\prime }nmm^{\prime
}}(q)=V_{nn^{\prime }m^{\prime }m}(q)=V_{n^{\prime }nm^{\prime }m}(q).
\end{equation}
On the other hand, the dispersion of the collective plasmon modes is given
by the zeros of the determinant of the dielectric tensor defined by Eq. (\ref
{diel22}), i.e., $\det \left| \varepsilon _{nn^{\prime }mm^{\prime
}}(q,\omega )\right| =0$. By using the symmetry properties of $V_{nn^{\prime
}mm^{\prime }}(q)$ into this determinant, one can show that the intersubband
plasmon modes are given by the roots of 
\begin{equation}
\varepsilon _{inter}=1-V_{1212}\left[ \Pi _{12}+\Pi _{21}\right] =0,
\label{detinter}
\end{equation}
whereas the intrasubband plasmon modes are obtained by the roots of 
\begin{equation}
\varepsilon _{intra}=\left[ 1-V_{1111}\Pi _{11}\right] \left[ 1-V_{2222}\Pi
_{22}\right] -V_{1122}^{2}\Pi _{11}\Pi _{22}=0.  \label{deintra}
\end{equation}
It is apparent that the intrasubband plasmon modes do not couple with the
intersubband ones. This is, of course, a direct result of our symmetric
confinement model which remains a reasonable model even in the presence of
small asymmetries in the 1D confinement.

First, we consider a high electron density ($N_{e}=N_{1}=10^{6}$ cm$^{-1}$)
in which case both the subbands are populated. The Fermi wavevectors in the
first and second subbands are $k_{F1}=1.09\times 10^{6}$ cm$^{-1}$ and $%
k_{F2}=0.47\times 10^{6}$cm$^{-1}$, respectively. In this case, two
intrasubband and two intersubband plasmon modes exist corresponding to the
two subbands. Fig. 1(a) shows the dispersion relations of the two
intersubband plasmon modes $(1,2)$ and $(1,2)^{\prime }$ obtained from Eq. (%
\ref{detinter}). The shadow areas indicate the intersubband single-particle
excitation (SPE$_{12}$) continua where $%
\mathop{\rm Im}%
\left\{ \Pi _{12}(q,\omega )\right\} \neq 0$. The intersubband SPE$_{12}$ is
of a finite frequency ($\omega =\omega _{0}$) at $q=0$. The occupation of
the second subband opens up a gap in the SPE$_{12}$ continuum where the
low-frequency intersubband plasmon mode $(1,2)^{\prime }$ appears. We also
see a large depolarization shift of the high-frequency intersubband plasmon
mode $(1,2)$. In Fig. 1(b) we show the dispersion relation of the
intrasubband plasmon modes $(1,1)$ and $(2,2)$ obtained from the Eq. (\ref
{deintra}). The intrasubband single-particle excitation SPE$_{11}$ (SPE$%
_{22} $) continuum where, $%
\mathop{\rm Im}%
\left\{ \Pi _{11}(q,\omega )\right\} \neq 0$ ($%
\mathop{\rm Im}%
\left\{ \Pi _{22}(q,\omega )\right\} \neq 0$), is also presented in the
figure. The undamped second subband intrasubband plasmon mode $(2,2),$ lying
in the gap between the SPE$_{11}$ and SPE$_{22}$ continua, has a linear
energy dispersion as $q\,\rightarrow 0$. When this plasmon mode enters the
SPE$_{11}$ continuum, it is Landau damped because it can decay by emitting
SPE$_{11}$ excitations in the lowest subband. The plasmon mode $(1,1)$
representing the collective charge-density excitation in the first subband
has an energy proportional to $q\left| \ln (qW)\right| ^{1/2}$. Notice that,
due to the symmetry of the system, the intersubband single-particle
excitations do not damp the intrasubband plasmon modes and vice versa. In
Fig. 2 we show the plasmon dispersion relations for a lower total electron
density $N_{e}=N_{2}=0.40\times 10^{6}$ cm$^{-1}$. In this case, only the
first subband is occupied. The corresponding 1D Fermi wavevector is $%
k_{F1}=0.63\times 10^{6}$ cm$^{-1}$. Obviously, the plasmon mode $(2,2)$ as
well as the continuum SPE$_{22}$ do not exist since the second subband is
empty. By analyzing the collective and single-particle excitation spectra in
Figs. 1 and 2 and comparing them with $\omega _{k}^{nn^{\prime }}(q)$
defined in Eq. (\ref{omega11}) we are able to figure out the contributions
of different scattering mechanisms to the total inelastic-scattering rate.

\subsection{Intrasubband scattering rate}

\smallskip According to Eq. (\ref{sigma}), the intrasubband inelastic
scattering rates $\sigma _{11}(k)$ and $\sigma _{22}(k)$ are defined in
terms of the imaginary part of the dynamically screened Coulomb potentials 
\begin{equation}
V_{1111}^{s}=\frac{V_{1111}\left( 1+V_{2222}\Pi _{22}\right)
-V_{1122}^{2}\Pi _{22}}{\varepsilon _{intra}}  \label{V1111s}
\end{equation}
and 
\begin{equation}
V_{2222}^{s}=\frac{V_{2222}\left( 1+V_{1111}\Pi _{11}\right)
-V_{1122}^{2}\Pi _{11}}{\varepsilon _{intra}},  \label{V2222s}
\end{equation}
respectively. These expressions are obtained from Eq. (\ref{diel}) and
demonstrate that the contributions to $\sigma _{11}(k)$ and $\sigma _{22}(k)$
come from three sources. The first one is the emission of the intrasubband
single-particle excitations. We numerically evaluate the integral in Eq. (%
\ref{sigma}) only in those regions where $%
\mathop{\rm Im}%
\Pi _{22}\neq 0$ $\,$($%
\mathop{\rm Im}%
\Pi _{11}\neq 0$). In Fig. 3(a) the thick-solid (thick-dashed) line shows
the intrasubband inelastic-scattering rate $\sigma _{11}(k)$ $\left(
\,\sigma _{22}(k)\right) $ due only to the emission of the single-particle
excitations in the SPE$_{22}$ (SPE$_{11}$) continuum. We take the impurity
broadening or the phenomenological damping constant $\gamma =10^{-3}$ meV
corresponding to samples with very high electron mobility. All other
parameters are the same as in the beginning of Sec. II. Our results show
that the intrasubband inelastic scattering in one subband takes place
through the emission of a single-particle excitation in the other subband.
In fact, we verified that $\omega _{k}^{nn}(q)$ defined in Eq. (\ref{omega11}%
) never crosses the SPE$_{nn}$ continuum in the $q$-$\omega $ plane for $n=1$
and $2$ independent of the value of $k$. This is the reason why the SPE$%
_{11} $ and SPE$_{22}$ continua do not contribute to $\sigma _{11}(k)\,$and $%
\,\sigma _{22}(k)$, respectively. We also verified that the curve $\omega
_{k}^{11}(q)$ is completely out of the SPE$_{22}$ continuum for $k\geq
k_{F2} $ and, consequently, the contribution of the SPE$_{22}\ $to $\sigma
_{11}(k)$ vanishes. On the other hand, the contribution of the SPE$_{11}\ $%
to $\sigma _{22}(k)$ starts at $k=k_{F1}$ where the curve $\omega
_{k_{F1}}^{22}(q)$ lies exactly on the lower edge of the SPE$_{11}$
continuum, entering the continuum for momentum $k>k_{F1}$. Thus, the SPE$%
_{22}$ contributes to $\sigma _{11}(k)$ from $k=0$ up to $k=k_{F2}$, while
the onset of the scattering $\sigma _{22}(k)$ via emission of a
single-particle excitation inside the continuum SPE$_{11}$ occurs at the
threshold $k=k_{F1}$. The thin-dashed line in Fig. 3(a) indicates the SPE$%
_{11}$ contribution to the scattering rate $\sigma _{22}(k)$ for an electron
in a quantum wire of a density $N_{e}=N_{2}$. At this density the second
subband is empty and as a consequence the SPE$_{22}$ contribution to $\sigma
_{11}(k)$ does not exist.

\smallskip According to Eqs. (\ref{V1111s}) and (\ref{V2222s}), the other
two sources contributing to $\sigma _{11}(k)$ and $\sigma _{22}(k)$ are the
two zeros of $\varepsilon _{intra}$, i.e. the emission of intrasubband
plasmon modes $(1,1)$ and $(2,2)$. In Fig. 3(b) we show the intrasubband
inelastic-scattering rate $\sigma _{nn}(k)$ due only to the intrasubband
plasmon modes $(n,n)$ with $n=1$ and $2.$ These results are obtained by
excluding the single particle excitation regions in the $q$-$\omega $ plane
where $%
\mathop{\rm Im}%
\left[ \Pi _{22}\right] \neq 0$ and $%
\mathop{\rm Im}%
\left[ \Pi _{11}\right] \neq 0$ from the numerical integration
characterizing $\sigma _{11}(k)$ and $\sigma _{22}(k)$, respectively. For $%
N_{e}=N_{1}$, the onset of both intrasubband scattering via emission of the
plasmon mode $(1,1)\,$occurs at the threshold $k=k_{c}^{11}\simeq 2.07\times
10^{6}$ cm$^{-1}$ corresponding to an interception of the curve $\omega
_{k}^{11}(q)$, as well as $\omega _{k}^{22}(q),$ with the mode $(1,1)$ at $%
q=q_{c}^{11}\simeq 0.55\times 10^{6}$ cm$^{-1}$. Indeed, the emission of the
plasmon mode $(1,1)$ is the most important contribution to the intrasubband
scattering rates due to its significant spectral weight at $q=q_{c}^{11}$
leading to a huge divergence at $k=k_{c}^{11}$. The onset of scattering due
to the plasmon mode $(2,2)$ occurs at the threshold $k=k_{F2}$ ($%
k_{c}^{22}\simeq 0.8\times 10^{6}$ cm$^{-1}$)\ for an electron in the first
(second) subband. Furthermore, the inelastic-scattering rate $\sigma
_{22}(k) $ at $k=k_{F1}$ is non-zero since $\omega _{k_{F1}}^{22}(q)$ lies
just on the lower edge of the SPE$_{11}$continuum where the plasmon $(2,2)$
contributes to scattering. But it no longer contributes to $\sigma
_{22}(k)\, $ for $k>k_{F1}$ due to the Landau damping. The fact that $\sigma
_{22}(k)$ at $k=k_{F1}$ is finite means that a plasmon mode $(2,2)$ may be
emitted when the electron has an energy $\xi _{2}(k_{F1})$. Finally, the
thin-solid (thin-dashed) line shows the contribution of the plasmon mode $%
(1,1)$ to $\sigma _{11}(k)$ ($\sigma _{22}(k)$) for $N_{e}=N_{2}$ where the
second subband is empty. As expected, the contribution coming from the
plasmon mode $(2,2)$ is absent in this case.

\subsection{Intersubband scattering rates}

\smallskip The definition in Eq. (\ref{sigma}) tells us that the
intersubband inelastic scattering rates $\sigma _{12}(k)$ and $\sigma
_{21}(k)$ are obtained in terms of the imaginary part of the screened
Coulomb potential 
\begin{equation}
V_{1221}^{s}=\frac{V_{1212}}{\varepsilon _{inter}}.  \label{VS1212}
\end{equation}
Notice that, according to Eq. (\ref{diel}), $V_{2112}^{s}=V_{1221}^{s}$.
Therefore the contributions to $\sigma _{12}(k),$ as well as $\sigma
_{21}(k),$ come from three sources: (i) the intersubband SPE$_{12}$
continuum; the intersubband plasmon modes (ii) $(1,2)$ and (iii) $%
(1,2)^{\prime }$. In Fig. 4(a) we show $\sigma _{12}(k)$ due only to the SPE$%
_{12}$ continuum. The thick and thin lines indicate the SPE$_{12}$
contributions to $\sigma _{12}(k)$ in the quantum wire of total charge
density $N_{e}=N_{1}$ and $N_{2}$, respectively. As discussed before, the SPE%
$_{12}$ continuum splits into two parts when the second subband is
populated. For $N_{e}=N_{1}$,\ the onset of scattering $\sigma _{12}(k)$ via
emission of an intersubband single-particle excitation in the lower (higher)
energy part of the SPE$_{12}\,$continuum occurs at $k\simeq 0.50\times
10^{6} $ cm$^{-1}$ ($k\simeq 1.46\times 10^{6}$ cm$^{-1}$). The most
important contribution of the lower part of the SPE$_{12}$ is shown in the
inset of Fig. 4(a) below the onset of the scattering the higher part.\ For $%
\sigma _{21}(k)$ we find no contribution of the single-particle excitations
due to restriction of the energy-momentum conservation defined by $\omega
_{k}^{21}(q)$. Therefore, electrons cannot transfer from the higher subband
to the lower one by emitting a single-particle excitation in the Fermi sea.

\smallskip In Fig. 4(b), the thick-solid (thick-dashed) line shows the
contributions to the inelastic-scattering rate $\sigma _{12}(k)$ $\left(
\sigma _{21}(k)\right) $ coming from the intersubband plasmon modes $(1,2)$
and $(1,2)^{\prime }$. Again, the thick\ and thin lines correspond to the
results for $N_{e}=N_{1}$ and $N_{2}$, respectively. Notice that the
intersubband plasmon modes do not contribute to $\sigma _{21}(k)$ for $%
k\lesssim k_{c}^{21}$, where $k_{c}^{21}\simeq 1.8\times 10^{6}$ cm$^{-1}$
and $1.31\times 10^{6}$ cm$^{-1}$ for electron density $N_{e}=N_{1}\,$\ and $%
N_{2}$, respectively. Because the curve $\omega _{k}^{21}(q)$ for $k$ $=$ $%
k_{c}^{21}$ never intercepts the mode $(1,2)^{\prime }$, the plasmon
contribution to the $\sigma _{21}(k)$ comes from the mode $(1,2)$. But both
the intersubband plasmon modes contribute to $\sigma _{12}(k)$. The onset of
scattering $\sigma _{12}(k)$ due to the emission of the plasmon mode $(1,2)$
occurs at the threshold $k_{c}^{12}\simeq 3.1\times 10^{6}$ cm$^{-1}$ ($%
2.61\times 10^{6}$ cm$^{-1}$) for $N_{e}=N_{1}\,$($N_{2}$). Indeed, we
verified that the curve $\omega _{k}^{12}(q)$ intercepts the plasmon mode $%
(1,2)$ for $k\geq k_{c}^{12}$ and the mode $(1,2)^{\prime }$ for all values
of $k$ except $k=k_{F1}$. As a matter of fact, we see the scattering rate $%
\sigma _{12}(k)$ vanishing at $k=k_{F1}$ due to restrictions of the
momentum-energy conservation dictated by the step functions in Eq. (\ref
{sigma}). In the inset, we show the most relevant contribution to $\sigma
_{12}(k)$ coming from the plasmon mode $(1,2)^{\prime }$.

We summarize the above results by plotting the total inelastic-scattering
rate 
\[
\sigma _{n}(k)=\sum_{n^{\prime }=1,2}\sigma _{nn^{\prime }}(k) 
\]
in Figs. 5 and 6, respectively, for an electron in the subband $n=1$ and $2$
in the quantum wire of total charge density $N_{e}=N_{1}$. The scale in the
right-hand side of Fig. 5 (Fig. 6) is enlarged 15 (5) times as compared to
that in the left. The symbols indicate the contributions from the different
scattering mechanisms. The circles stand for the contribution coming from
the emission of a plasmon mode $(2,2)$, while the squares represent the
contribution coming from the emission of the plasmon mode $(1,1)$. The
filled (open) triangles-left stand for the contribution coming from the
emission of a intersubband plasmon mode $(1,2)$ $\left( (1,2)^{\prime
}\right) .$ The open (filled) diamonds represent the contribution coming
from single-particle excitations inside the SPE$_{22}$ $\left( \text{SPE}%
_{11}\right) $ continuum. Finally, open (filled) triangles-up in Fig. 5
represent the contribution coming from the single-particle excitations in
the lower (higher) part of the SPE$_{12}$ continuum. These symbols show the
complexity of various intra- and inter-subband single particle and
collective mode contributions to the scattering rate of an electron which
might be scattered either to unoccupied states in its original subband or to
those in a different subband. In contrast to $\sigma _{2}(k)$, we see $%
\sigma _{1}(k)$ being finite at $k=0$ due to the possibility of emission of
a single-particle excitation within the SPE$_{22}$ continuum. As $k$
increases, the plasmon mode $(1,2)^{\prime }$ starts to contribute to $%
\sigma _{1}(k)$ and then, as we discussed above, all excitations in the
phase space contribute to scattering. In Fig. 5, the open triangles-up
indicate the contribution coming from the single-particle excitations inside
the lower energy part of the SPE$_{12}$ continuum. Notice that we neglected
such a contribution in the right-hand side since it is irrelevant at that
scale. When the second subband is empty for $N_{e}=$ $N_{2}$, we should see
neither the contributions coming from the plasmon modes $(1,2)^{\prime }$
and $(2,2)$ nor those from the SPE$_{22}$ continuum.

\subsection{Hot electron lifetimes}

We now discuss the hot electron lifetime 
\begin{equation}
\tau _{E,n}=\frac{2}{\sigma _{n}(k)}  \label{tau}
\end{equation}
of an energetic hot electron injected in a subband $n$ with a kinetic energy 
$E=\hbar ^{2}k^{2}/2m^{*}$ above the Fermi energy $E_{Fn}=\hbar
^{2}k_{n}^{2}/2m^{*}$ in the subband. It is well-known that this lifetime ($%
\hbar =1$) can be written as Eq. (\ref{tau}) since $\sigma _{n}(k)$ is the
absolute value of the imaginary part of the self-energy of an electron in
the subband $n$. \cite{mhbook} In Figs. 7(a) and 7(b) we show the lifetimes $%
\tau _{E,1}$ and $\tau _{E,2}$ of hot electrons injected in the first and
the second subband, respectively, as a function of the total electron
density $N_{e}$ in the quantum wire. The vertical thin line at $%
N_{e}=0.63\times 10^{6}$cm$^{-1}$ indicates the onset of the population of
the second subband for our quantum wire parameters. The thick-solid
(thick-dashed) lines indicates the lifetime of an injected ''hot electron''
in the conduction band with total energy $E_{T1}=E_{2}+\omega _{0}$ $\left(
E_{T2}=E_{2}+3\omega _{0}\right) $, where $E_{2}$ is the bottom of the
second subband. The symbols stand for the same meaning as in Figs. 5 and 6,
and show the contributions of the different charge-density excitations to
the total lifetime. We see the plasmon mode $(1,1)$ making the most
important contribution (solid squares) to $\tau _{E,1}\,$ for $E=E_{T1}$
(thick solid lines) at low densities. The decreasing of $\tau _{E_{T1},1}$
as $N_{e}$ increases indicates the hot-electron relaxation via emission of a
plasmon mode $(1,1)$. The single-particle excitations inside the SPE$_{12}$
continuum (filled triangles-up) start to contribute to $\tau _{E_{T1},1}$ as
the onset of the scattering via emission of the plasmon mode $(1,1)$
vanishes at $N_{e}\gtrsim 0.35\times 10^{6}$cm$^{-1}$. For $E=E_{T2}$,
however, the SPE$_{12}$ is the main contribution to $\tau _{E,1}$ at very
low densities. With increasing density, scattering due to the plasmon mode $%
(1,1)\ $becomes dominant until $N_{e}\simeq 0.83\times 10^{6}$ cm$^{-1}$.
The occupation of the second subband only leads to a small contribution by
the intersubband plasmon $(1,2)^{\prime }$ (triangles-left) to $\tau _{E,1}$%
. Although the onset of the scattering by the plasmon mode $(1,2)$ is
achieved at low density $N_{e}$, its contribution to $\tau _{E,1}$ is so
small that we cannot observe it in the figure. The contribution coming from
the SPE$_{22}$ is also irrelevant in this situation. As discussed before,
this excitation exists as the second subband is occupied but its
contribution to scattering vanishes for momentum $k\geqslant $ $k_{F2}.$
Here, we are dealing with a hot-electron of energy $E\gg \left( \hbar
k_{F2}\right) ^{2}/2m^{*}$ for which relaxation via the emission of
single-particle excitations inside the SPE$_{22}$ continuum is not allowed.

In Fig. 7(b), we see that the emission of the plasmon mode $(1,2)$ (filled
triangles-left), at very low densities, is the main contribution to the
total lifetime $\tau _{E,2}\,$for both values of $E$. As $N_{e}$ increases,
the hot-electron scattering via emission of the plasmon mode $(1,1),$ as
well as the emission of single-particle excitations inside SPE$_{11}\ $%
continuum (filled diamonds), start to contribute to $\tau _{E,2}$. For
densities greater than $N_{e}\simeq 0.36\times 10^{6}$ cm$^{-1}$ ($%
N_{e}\simeq 0.44\times 10^{6}$ cm$^{-1}$) the onset of scattering via the
emission of the plasmon mode $(1,1)$ $\left( (1,2)\right) $ vanishes, so
that only single-particle excitations inside the SPE$_{11}$ continuum
(filled diamonds) are responsible for the hot-electron relaxation in the
second subband. For $E=E_{T2}$, the onset of scattering via the emission of
the plasmon mode $(1,2)$ occurs at densities $N_{e}>10^{6}$ cm$^{-1}$. As a
result, this mode contributes to $\tau _{E,2}$ for all values of $N_{e}\,$%
shown in the figure.

At this point, we should briefly comment on the role of the phenomenological
damping constant $\gamma $ on our numerical results since we have used $%
\gamma =10^{-3}$ meV throughout this paper. In contrast to Figs. 5 and 6,
where effects of $\gamma $ ($=10^{-3}$) are vanishingly small, a finite $%
\gamma $ has some effect in Fig. 7. Notice that, for extremely clean systems
($\gamma =0$), the contributions to both $\tau _{E,n},$ coming from the
plasmon mode $(1,1)$ (square lines), should go to infinity threshold due to
the singular nature of 1D density of states. Impurity scattering through a
finite $\gamma $ suppresses this divergence by smoothing the 1D density of
states. Similar behavior should occur for the contribution to $\tau _{E,2}$
coming from the emission of the mode $(1,2)$ (filled triangles-left) in Fig.
7(b). In fact, effects due to finite values of $\gamma $ on both
hot-electron lifetimes can be identified in the lines indicating the
contribution of the plasmon modes $(1,1)$ and $(1,2)$. They do not go to
infinity since we are using $\gamma =10^{-3}$ meV, which is enough to
suppress the 1D density of states singularity on the scales of these
figures. We see, however, that these effects are irrelevant for the total
hot-electron lifetime when the emission of single-particle excitations are
taken into account. For the value of $\gamma $ assumed in this paper, the
relaxation of hot electrons is mainly due to emission of charge-density
excitations in the Q1D Fermi sea.

\section{Summary}

Within the GW approximation, we have calculated the inelastic Coulomb
scattering rates and lifetimes of an injected electron in a symmetric
confinement two-subband quantum wire at zero temperature. These rates are
directly related to the dynamically screened Coulomb potential which has
been calculated within the RPA. We chose a quantum wire with symmetric
confinement potential in which the intra- and intersubband excitations do
not interact with each other. We obtain the effects of the population of
second subband on the inelastic Coulomb scattering rate. We separately
identified the contributions to the intrasubband and intersubband
inelastic-scattering rates due to different intrasubband and intersubband
excitations in the individual subbands of the Q1D electron system. We find
the emission of an intrasubband plasmon in the first subband to be the most
important contribution to the inelastic-scattering rate, although the
single-particle excitations as well as plasmon modes in the second subband
also contribute to the intrasubband and intersubband scattering of an
electron in the two-subband quantum wire. We found that the inelastic
scattering from the first to the second subband occurs through the emission
of either intersubband plasmon modes or intersubband single-particle
excitations, whereas the scattering from second to first subband only occurs
via the emission of the higher energy intersubband plasmon mode. We also
calculate the lifetime of hot electrons as a function of the total charge
density in the two-subband quantum wire, identifying the contributions of
plasmons and single particle excitations in each subband to the hot electron
lifetime.

\section*{Acknowledgments}

We would like to thank E.H. Hwang for very useful discussions. The work of
MRST is supported by {\em FAPESP}, Brazil. The work at Maryland is supported
by the US-ARO and US-ONR. G-QH acknowledgs {\em CNPq} from Brazil for
partial support.

\begin{figure}[tbp]
\caption{ Dispersion relation of both (a) inter- and (b) intrasubband
charge-density excitations in a quantum wire of N$_e$=N$_1$=10$^6$ cm$^{-1}$
with $\omega_0$=5.37 meV. The shadow areas indicate the (a) inter- and (b)
intrasubband single-particle continua. Both subbands are occupied.}
\end{figure}

\begin{figure}[tbp]
\caption{ Dispersion relation of both inter- and intrasubband charge-density
excitations in the same quantum wire of N$_e$=N$_2$=0.4 $\times$ 10$^6$ cm$%
^{-1}$. The shadow areas indicate the (a) intersubband and (b) intrasubband
single-particle continua. Only the lowest subband is occupied}
\end{figure}

\begin{figure}[tbp]
\caption{ Intrasubband inelastic-scattering rates $\sigma_{11}$ (solid
lines) and $\sigma_{22}$ (dashed lines) due to emission of intrasubband (a)
single-particle and (b) collective excitations. Thick (thin) lines represent
results for our quantum wire of density N$_e$=N$_1$ (N$_e$=N$_2$). }
\end{figure}

\begin{figure}[tbp]
\caption{Intersubband inelastic-scattering rates $\sigma_{12}$ (solid lines)
and $\sigma_{21}$ (dashed lines) due to intersubband (a) single-particle and
(b) collective exciatations. Thick (thin) lines represent results for a
quantum wire of density N$_e$=N$_1$ (N$_e$=N$_2$). Inset in part (a) shows
the $\sigma_{12}$ due to only those single-particle excitations in the lower
energy part of the SPE$_{12}$ continuum. Inset in part (b) shows $%
\sigma_{12} $ only due to the plasmon mode (1,2)$^{\prime}$. }
\end{figure}

\begin{figure}[tbp]
\caption{Inelastic-scattering rate $\sigma_n(k)$ of electrons in the first
subband (n=1). The density in the quantum wire is N$_e$=N$_1$. }
\end{figure}

\begin{figure}[tbp]
\caption{Inelastic-scattering rate $\sigma_n(k)$ of electrons in the second
subband (n=2). The density in the quantum wire is N$_e$=N$_1$. }
\end{figure}

\begin{figure}[tbp]
\caption{Hot electron lifetimes (a) $\tau_{E,1}$ and (b) $\tau_{E,2}$ as a
function of the total density $N_e$ in the quantum wire. Thick solid
(dashed) lines are the total lifetime of a hot electron with $E=E_{T1}$ ($%
E=E_{T2}$). The thin lines indicate all sort of contribution to the total
lifetime. The symbols stand for the same contributions as in Figs. 5 and 6. }
\end{figure}

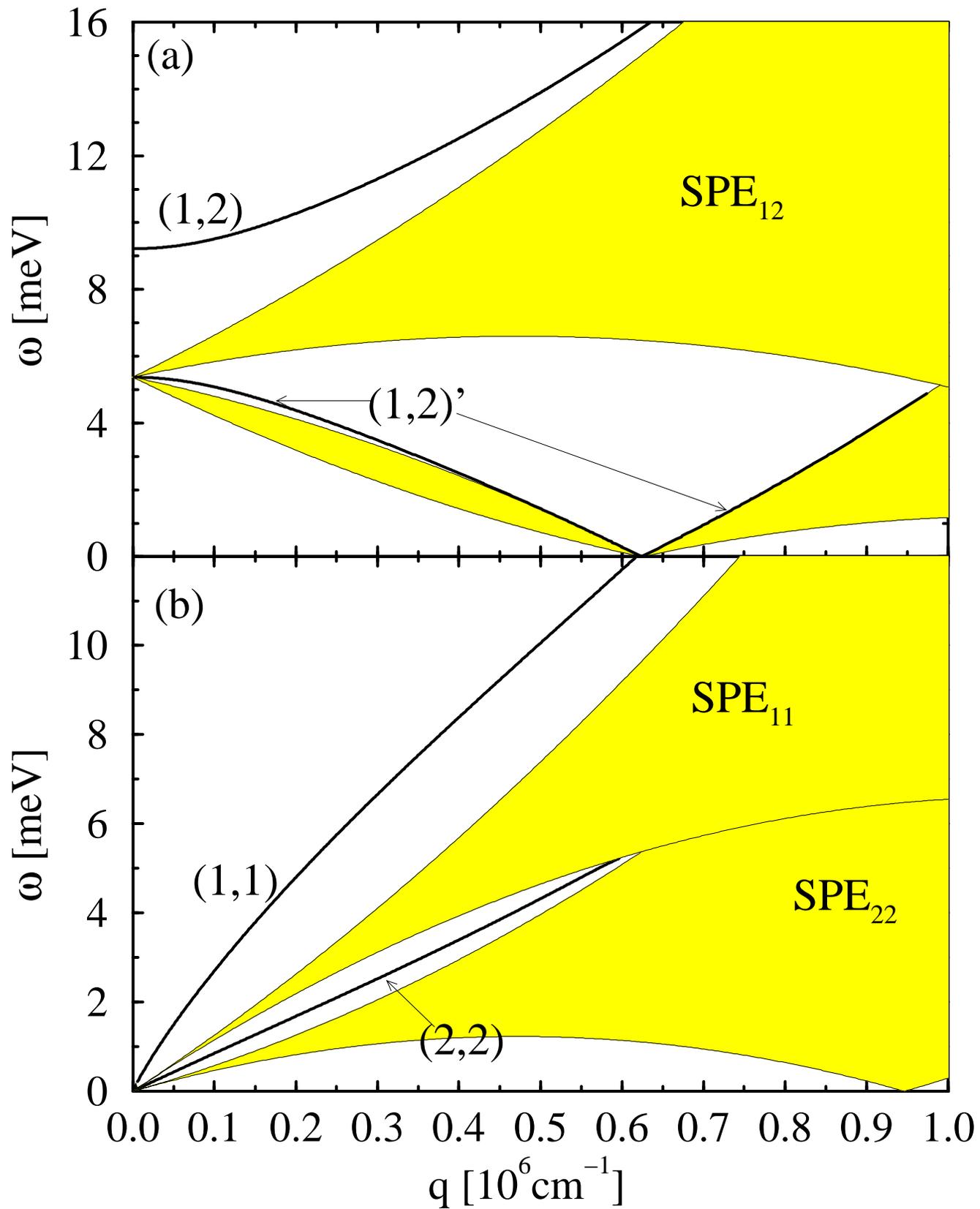
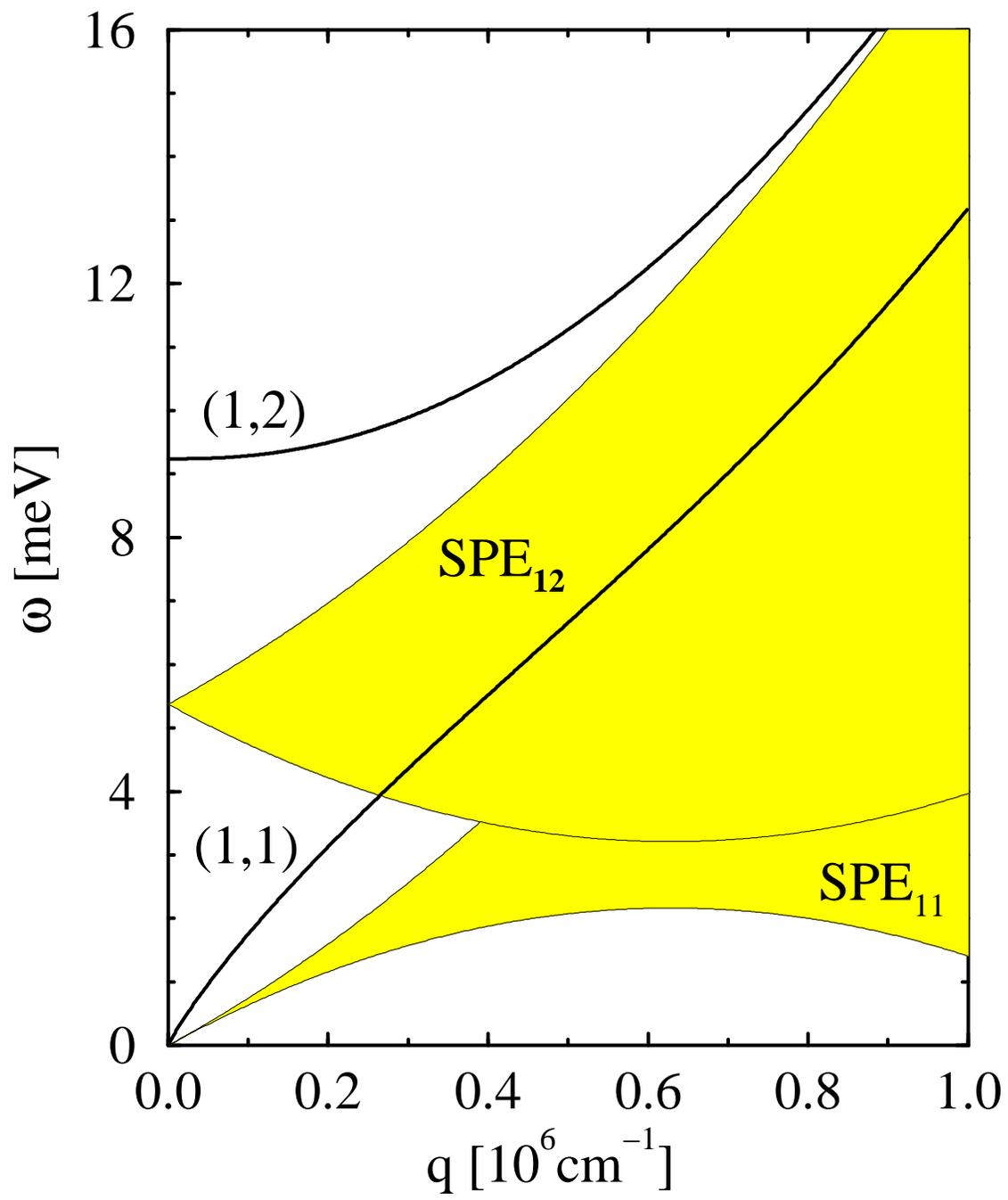
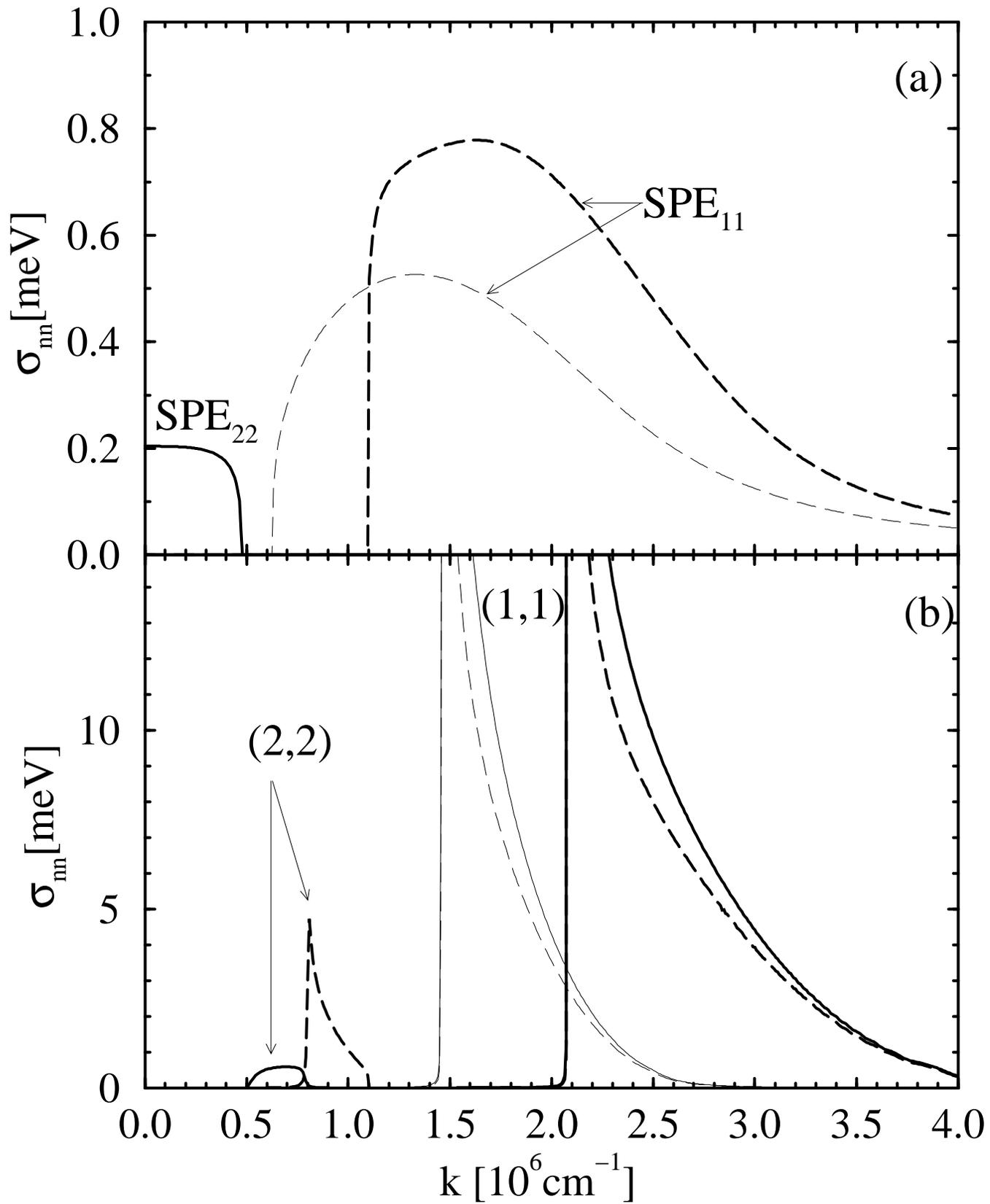
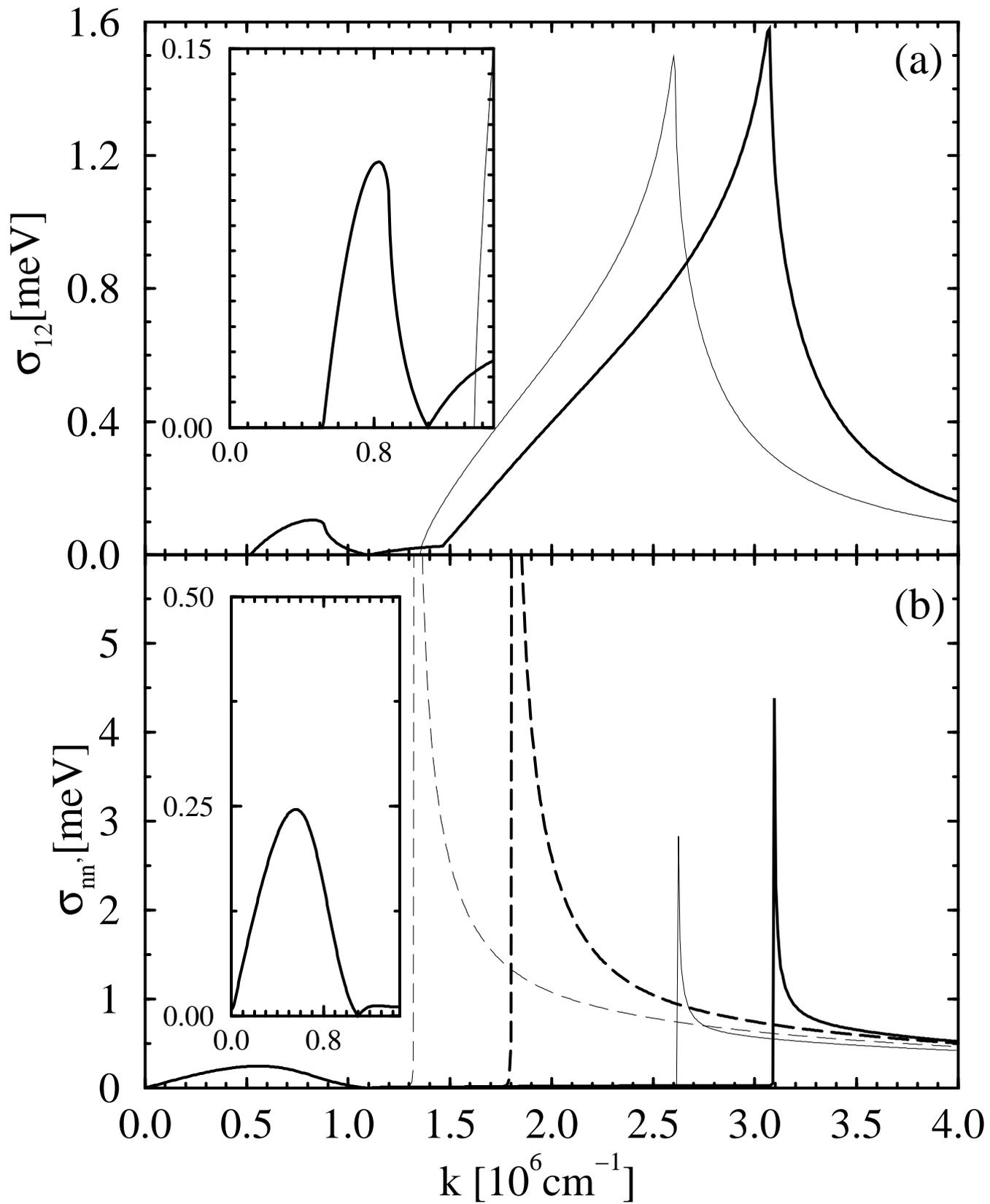
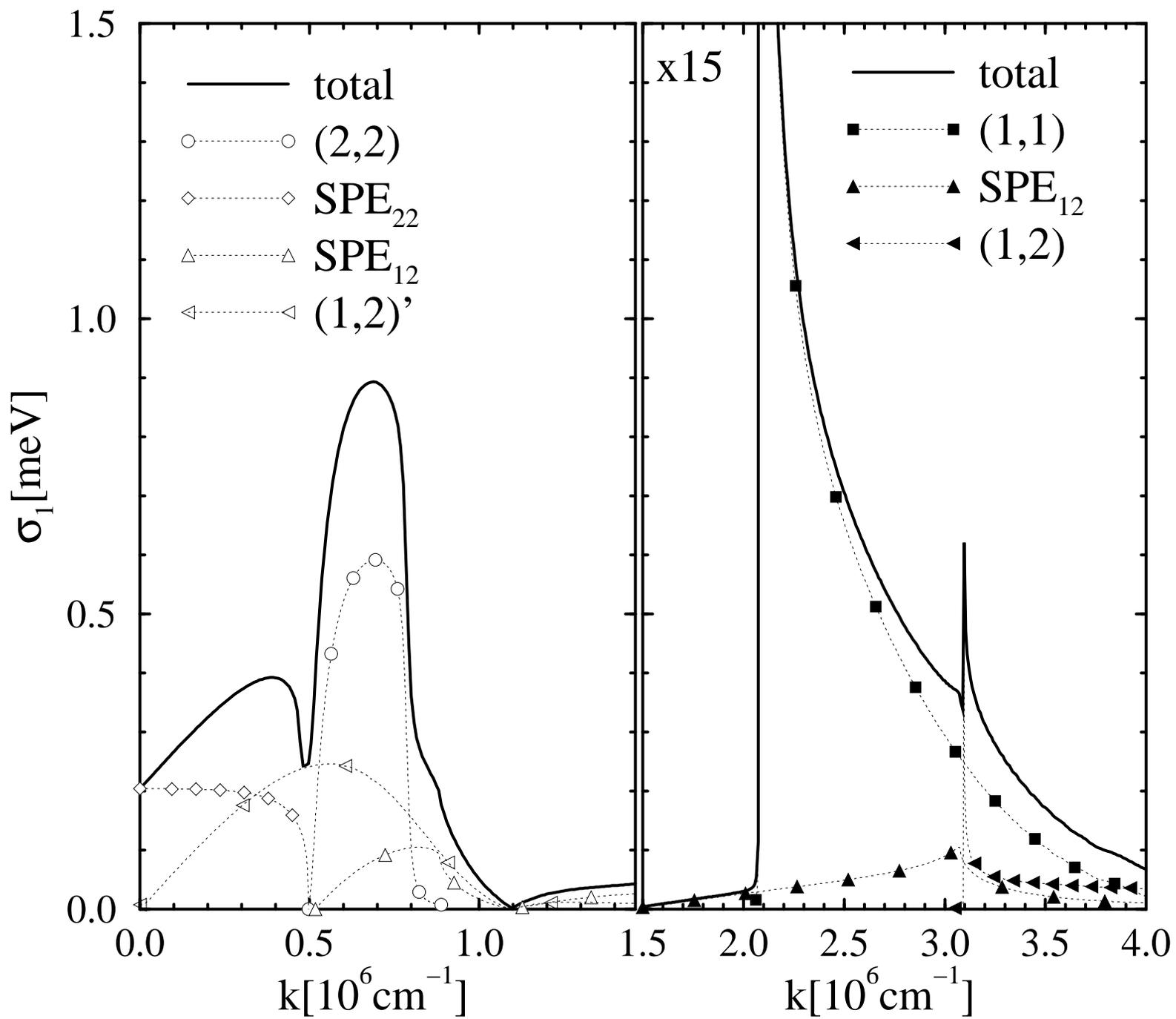
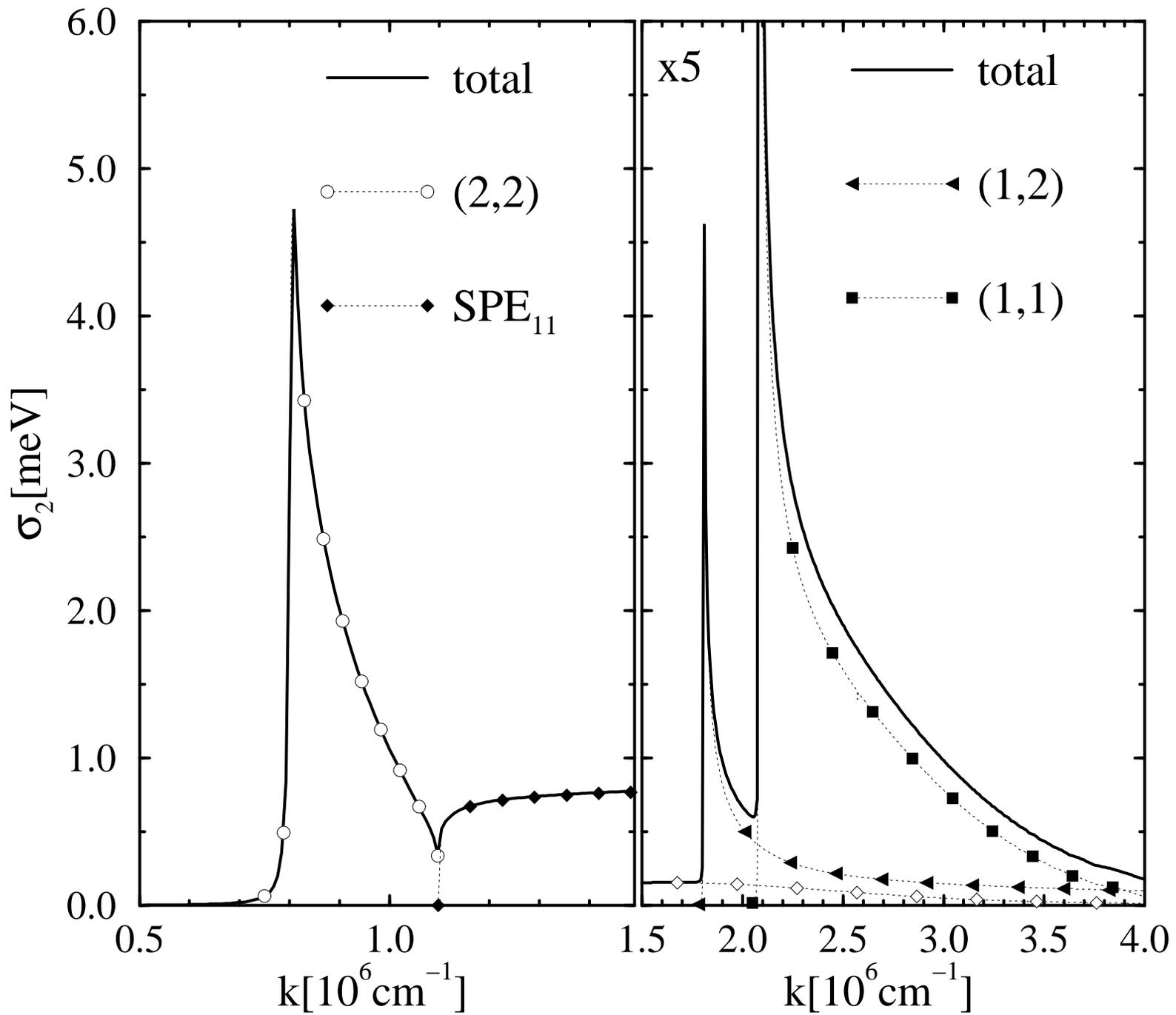
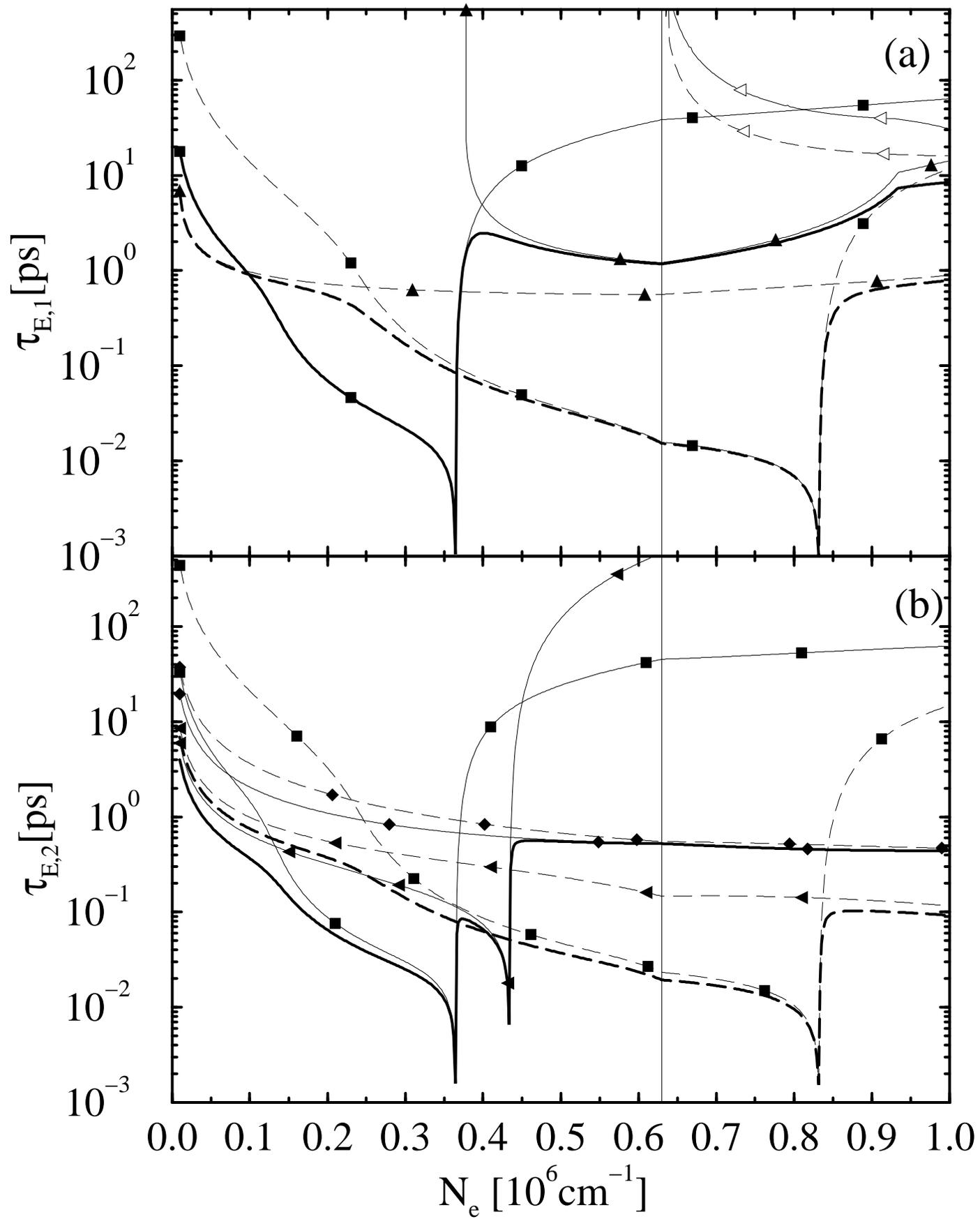
\end{document}